# Spectral Similarity Masks Structural Diversity at Hydrophobic Water Interfaces

Yong Wang[1], Yifan Li[1], Linhan Du[2], Chunyi Zhang[1], Lorenzo Agosta[1,3], Marcos Calegari Andrade[4,5], Annabella Selloni[1]*, and Roberto Car[1]*

[1] *Department of Chemistry, Princeton University, Princeton, New Jersey 08544, USA*

[2] *Department of Chemical Engineering, Tsinghua University, Beijing 100084, China*

[3] *Department of Chemistry, Ångström Laboratory, Uppsala University, 751 21 Uppsala, Sweden*

[4] *Chemistry and Biochemistry Department, University of California, Santa Cruz, California, 95064, USA*

[5] *Quantum Simulations Group, Materials Science Division, Lawrence Livermore National Laboratory, Livermore, California, 94550, USA*

*Emails: aselloni@princeton.edu, rcar@princeton.edu,*

**ABSTRACT:**

The air-water and graphene-water interfaces represent quintessential examples of the liquid-gas and liquid-solid boundaries, respectively. While the sum-frequency generation (SFG) spectra of these interfaces exhibit certain similarities, a consensus on their signals and interpretations has yet to be reached. Leveraging deep learning, we accessed fully first-principles SFG spectra for both systems, addressing recent experimental discrepancies. Despite both interfaces exhibiting microscopically hydrophobic characteristics, our findings reveal that similarities in SFG signals do not translate into comparable interfacial microscopic properties. Instead, graphene-water and air-water interfaces exhibit fundamental differences in SFG-active thicknesses, hydrogen-bonding networks, and dynamic diffusion behavior. These distinctions underscore the stronger confinements imposed by the solid-liquid interface compared with the weaker constraints of the gas-liquid interface.

**INTRODUCTION**

The interfaces between liquid water and gas/solid phases play a central role in natural processes and technological applications[1]. For example, the air-water interface governs atmospheric processes such as evaporation, condensation, and gas exchange[2], while the graphene–water interface provides a convenient substrate for catalysis, adsorption, and biomolecular functions[3,4]. Moreover, the graphene–water interface serves as a critical platform for modulating a wide range of phenomena in confined environments[5,6]. A molecular-level understanding of these interfaces is therefore essential for tackling energy and environmental challenges.

Vibrational sum frequency generation (vSFG) spectroscopy, with its inherent surface sensitivity, has emerged as a powerful probe of interfacial water[7,8]. Recent substrate-free heterodyne-detected SFG (HD-SFG) experiments reveal notable similarities between graphene–water and air-water interfaces. Bonn et al.[9] (E1) observed a broad negative feature around 3400 cm$^{-1}$ attributed to hydrogen-bonded oxygen-hydrogen (OH) groups pointing toward the bulk and a positive peak near 3650 cm$^{-1}$ attributed to free OH groups facing air or graphene, with the graphene–water interfacial feature showing a 30±10 cm$^{-1}$ redshift and intensity variations indicative of weak interactions. Tian et al.[10] (E2) reported a larger redshift (~100 cm$^{-1}$) and a 50% reduction in the amplitude of HD-SFG compared to air-water, with significant contributions from graphene and bulk water. Other studies show that the positive peak in the HD-SFG signal of the graphene-water interface even vanishes when the graphene layer is supported by a substrate[11,12], illustrating the strong sensitivity of HD-SFG to interfacial modifications.

Computational studies offer molecular-scale insights that complement experiments. Molecular dynamics (MD)-based calculations of surface-specific vibrational SFG spectra have validated experimental observations and elucidated the interfacial structure[13–16]. However, accurate MD simulations require precise interatomic potentials capable of modeling complex interactions that empirical force fields often do not capture[8]. *Ab initio* MD (AIMD)[17–19] can calculate the resonant SFG response $\chi^{(2)}(\omega)$ from the Fourier transform of the dipole moment–polarizability ($\mu - \alpha$) time correlation function, but well-converged calculations can require nanosecond-scale simulations[20]. Alternatively, computational cost can be substantially reduced by approximations such as the surface-specific velocity–velocity correlation function (ssVVCF) method[15], which focuses on the OH stretching mode and only requires simulations spanning tens of picoseconds[21]. Recent advances in machine learning (ML) for MD[22–25] have further facilitated the calculation of vibrational spectra. ML-based MD (MLMD), trained with small-scale *ab initio* data, makes simulations of larger systems and longer timescales possible, allowing efficient calculations of the SFG spectra with the ssVVCF method[9,26]. In addition, ML models have successfully predicted the dipole moments and polarizability tensors of dynamic environments[25,27–29], opening the way to fully first-principles SFG calculations for interfaces such as air–water[30–32], air–ice[33], and $\alpha$-$Al_2O_3$(0001)–water[21]. In parallel with these advances, recent simulations that utilize the MB-pol potential for water have provided deeper insights into the influence of surface charge on the SFG signals of graphene–water interfaces[34].

In this work, we use deep learning techniques[23,27,28] to compute from first principles with the $\mu - \alpha$ method the HD-SFG spectra of the air–water and graphene–water interfaces. Our results are in excellent agreement with the E1 experiment[9] and help better understand the different observations made in other experiments[9–12]. Our investigations confirm the remarkable spectral similarity between the air-water and graphene-water interfaces, but they also display significant differences between the two interfaces in the SFG-active thickness, the evolution of the hydrogen-bond network, and the diffusion dynamics. Although both interfaces share a hydrophobic character at the molecular scale, these differences arise from the distinct nature of the solid-liquid and gas-liquid boundaries, where the former imposes a stronger confinement on the interfacial water molecules than the latter.

**RESULTS AND DISCUSSION**

**Structural features of the air-water-graphene slab**

To characterize air-water and graphene-water interfaces, we constructed three machine learning models based on *ab initio* data using the SCAN exchange-correlation functional[35]. Comparisons with other functionals can be found in the supplementary information. The three models describe the interatomic potential interaction, the molecular dipole moments, and the polarizability of liquid water. The machine learning potential (MLP) enables large-scale molecular dynamics simulations, while the dipole and polarizability models allow efficient on-the-fly computation of the $\mu - \alpha$ correlation. Figure 1a presents a side view snapshot of the graphene-water-air slab from MLMD simulations, where interfacial water exhibits OH groups not participating in H bonds oriented toward both interfaces. The simulation cell contains 60 carbon atoms in the graphene monolayer, 166 water molecules, and a vacuum slab with dimensions of 12.01 Å×12.78 Å in the $xy$ plane and 50 Å along z, including a 15 Å vacuum layer to mitigate periodic boundary effects[8]. This slab model contains two interfaces and allows water to equilibrate more freely without constraint compared to the confined models[34,36]. Twelve independent MLMD simulations were performed in the canonical ensemble for 1 ns at 340 K to reduce statistical errors, with about 40 K elevation to mitigate the SCAN functional's overestimation of ice's melting temperature[37], with molecular dipole and polarizability computed on-the-fly. Figure 1b shows the number density of carbon atoms (as the z-axis physical density is not well-defined in monolayer graphene) and water density along z in the MLMDs. The bulk water region exhibits a stable density of 1.04 $g/cm^3$, slightly exceeding the experimental value, likely due to the SCAN functional's tendency to overestimate the strength of the H-bonds[38]. Figure 1c shows the dipole moment distribution along $z$, distinguishing the graphene-water and air-water interfaces from the bulk. The average dipole moment in liquid water is 3.05 Debye, higher than the gas-phase value of 1.85 Debye. At the graphene-water interface, the dipole moment decreases to about 2.5 Debye, with notable variations limited to the first two layers. Figure 1d presents the $\alpha_{yy}$ component of the polarizability tensor along z (the different directions in our model are illustrated in Supplementary Fig. 2). At the air-water interface, $\alpha_{yy}$ decreases from 1.76 Å³ in the liquid to 1.45 Å³ in the gas, while at the graphene-water interface, variations are small and confined to the first two layers. These findings suggest that interfacial density variations alone do not dictate dipole and polarizability behavior. The machine-learning models predict water properties in agreement with experiments, including dipole moments in the gas[39] and liquid phase[40], as well as gas-phase polarizability[41].

To facilitate further analyses, we divided both interfaces into three layers. The graphene-water interface is partitioned based on water density fluctuations, while for the air-water interface, the first layer defines the Gibbs dividing surface (where surface excess is zero, very near the point where the water density equals half its bulk value), the second marks the onset of polarizability changes, and the third is a 2.5 Å-thick slab at its inner edge next to the bulk. This division enables a comparative analysis of the two interfaces, although alternative partitioning methods exist[42]. Figures 1e and 1f show the distribution of the angle $\theta$ between the water dipole vector and the surface normal across different interfacial layers and the bulk for both air–water and graphene–water interfaces. At the air-water interface, the third layer and the bulk exhibit an isotropic angular distribution following $sin\theta$, while the first and second layers show pronounced peaks around 90°, with stronger alignment in the first layer. This confirms the presence of dangling OH bonds deviating from the ideal alignment ($\theta = 52.25°$) with the surface normal. At the graphene–water interface, only the first layer ($z \leq 5.3\,\text{Å}$) deviates significantly from isotropic behavior, displaying a bimodal angular distribution, consistent with Ref.[34]. This distribution can be further decomposed into contributions from water near graphene ($z < 4\,\text{Å}$) and slightly further away ($4\,\text{Å} \leq z \leq 5.3\,\text{Å}$). Water molecules closer to graphene exhibit OH dangling bonds more oriented toward the interface than in the air-water case, as indicated by the leftward shift in $\theta$, due to the weak attraction between OH bonds and graphene. Notably, small water density shoulders at the graphene-water interface ($z \sim 4\,\text{Å}$) define these orientation regions. This feature likely arises from SCAN inducing greater interfacial ordering compared to functionals such as revPBE-D3 (Supplementary Fig. 11). However, this does not affect our main conclusions, as revPBE-D3 exhibits similar interfacial trends (Supplementary Fig. 11).

**vSFGs of the graphene-water and the air-water interfaces**

The accuracy of our machine-learning models enables first-principles calculations of HD-SFG spectra for air-water and graphene-water interfaces. To enable direct comparison with experimental results, we use a scaling factor of 0.96 along the frequency axis of the calculated spectra, compensating for the nuclear quantum effects in the high-frequency region, as employed in previous studies[8]. Figure 2 (Top) compares our $\mu - \alpha$ approach with two recent substrate-free experiments (E1, E2)[9,10]. Our results align well with E1, showing: (1) A relative peak intensity ratio close to 1 between the two interfaces, with the calculated and experimental air-water high-frequency peaks scaled to 1 and the graphene-water signals adjusted accordingly; (2) A 41 cm$^{-1}$ redshift of the graphene-water high-frequency positive peak relative to the air-water interface, attributed to weak interactions between "dangling" OH bonds and the conjugated graphene $\pi$-bond, consistent with Fig. 1f; (3) Peak frequencies of the high-frequency positive mode matching experimental observations. Our results also capture the shoulder of the positive peak in the air-water signal at ∼3600 cm$^{-1}$ (experimentally at ∼3650 cm$^{-1}$), which is absent in ssVVCF-based SCAN calculations[43]. This discrepancy has been linked to omitting environment-dependent dielectric responses in the velocity correlation method[32]. However, the low-frequency negative feature in our calculations is broader than in experiments, due to the SCAN's known tendency to overestimate feature widths[43]. Moreover, in E1 there is a deeper minimum for graphene-water than for air-water, and the tail is red-shifted (slightly) at graphene-water. These trends are reproduced by our calculations but are enhanced relative to the experiment. Previous ssVVCF calculations[9] showed a similar low-frequency tail with the BLYP+D3 functional but not with the revPBE+D3 functional. These findings highlight the future need to compare experimental setups and clarify spectral features by considering quadrupole and non-resonant contributions[8].

Building on the agreement between the total computed HD-SFG signal and experimental results, decomposing the signal into contributions from different interfacial layers and hydrogen-bonding (H-bond) species provides valuable molecular-level insights into the microscopic origins of the observed signals. Figure 2 (middle) shows the decomposition of the signals into three layers at the two interfaces, as defined in Figure 1b. At the air-water interface, the first two layers dominate the signal, where density, dipole, and polarizability decrease continuously. The two layers show similar spectral trends, with the first being slightly weaker due to fewer molecules and the second slightly red-shifted owing to its proximity to the bulk. This is consistent with the change of molecular orientation at the air-water interface (Fig. 1e). The molecular properties and signal decomposition collectively illustrate that the transition of water from liquid to gas at the air-water interface is continuous and smooth, without a strong transition of structural order. This arises from the weak constraint of air at the interface, allowing for greater flexibility in interfacial changes. For the graphene-water interface, signal decomposition into different layers reveals a stark contrast to the air-water interface. Nearly the entire SFG signal originates from the topmost, first layer, while contributions from the second and third layers abruptly drop to zero without a gradual decrease. This behavior aligns with Fig.1f and the findings of E1, which showed that the contribution of bulk water—both neat and electrolyte—to the total HD-SFG signal is negligible[9]. Thus, we define an interfacial depth, $d_\chi$, to represent the water thickness over which the full SFG signal can be contributed. In our results, $d_\chi^{air} \approx 9.5$ Å and $d_\chi^{graphene} \approx 2.8$ Å.

Figure 2 (bottom) depicts the decomposition of total HD-SFG signals at the air-water and graphene-water interfaces into contributions from water molecules with distinct hydrogen-bond (H-bond) characteristics. H-bond formation is defined using geometric criteria: an O···O distance < 3.5 Å and an O-H···O angle > 130°. Following previous studies[8,9,21], water molecules were classified into three groups based on the number of donated ($n_D$) and accepted ($n_A$) H-bonds: DAA+DA ($n_D \leq 1$), DDA ($n_D > 1, n_A \leq 1$), and DDAA ($n_D > 1, n_A > 1$). Here, "D" and "A" represent H-bond donors and acceptors, respectively. The DA and DAA categories were combined into DAA+DA due to their free O-H bonds, which are essential for spectral decomposition. Minor species such as A, D, DAAA, DDDA, DDAAA, and DDDAA were included within their respective categories. Due to thermodynamic fluctuations, a small fraction of DAA+DA and DDA species can be found within the bulk region. Despite simplifying the complex H-bond network at the interfaces, this classification retains essential details for spectral decomposition (Fig. 3a). The high-frequency positive peaks observed at both interfaces primarily originate from water molecules in the DAA+DA configurations, which donate a single hydrogen bond and orient their remaining free O–H bond toward the air or graphene side of the interface. In addition to this contribution, the DAA+DA group is also involved in the broad negative feature, as their hydrogen-bonded O–H group points toward neighboring molecules in the bulk, reflecting their asymmetric interfacial orientation. For DDA species, their contributions are primarily concentrated on the right side of the negative feature at both the air-water and graphene-water interfaces, with slightly lower intensity at the air-water interface. The DDAA group shows distinct spectral behavior at the two interfaces. At the air-water interface, DDAA contributes only a minor portion within the negative spectral region. In contrast, at the graphene-water interface, DDAA contributes a notable positive peak within the same negative spectral region. This highlights the ordering differences in the H-bond networks at the two interfaces, particularly in the transition region from the interface to the bulk, where DDAA is concentrated (Fig. 3a). These behaviors arise from the distinct interfacial constraints: the air-water interface ($d_\chi^{air} \approx 9.5\,\text{Å}$) allows a gradual transition and less ordering in molecular structure, while the rigid graphene interface ($d_\chi^{graphene} \approx 2.8\,\text{Å}$) imposes strong confinement, compressing interfacial non-centrosymmetric regions and hydrogen bond networks (Fig. 3a).

**Evolution of H-bond, tetrahedral order, and diffusion**

The decomposition of the total computed HD-SFG spectra reveals subtle differences between the graphene-water and air-water interfaces, despite their overall similarity in total SFG signals. In the liquid phase, water molecules form a local tetrahedral structure due to directional hydrogen bonding, where each water molecule typically bonds with four neighbors. To further examine the topology of the hydrogen-bond network at the interfaces, Figure 3b presents the evolution of the nearest-neighbor water count $N$ and the tetrahedral order parameter $q$ at both interfaces, using a cutoff radius $r_c = 3.15\,\text{Å}$. The order parameter $q$ is defined as

$$q = \begin{cases} 1 - c_N \sum_{j=1}^{N-1} \sum_{k=j+1}^{N} \left(\cos \psi_{jk} + \frac{1}{3}\right)^2, & \text{if } N \geq 2, \\ 0, & \text{if } N = 0,1, \end{cases}$$

$$c_N = \frac{9}{2N(N-1)},$$

accounting for reduced coordination at the interface. At the graphene-water interface, the average $N$ decreases from about 4 to about 2.5, while at the air-water interface, it drops to 0, reflecting gas-phase behavior. In the bulk, $q$ averages 0.72, consistent with previous work[44]. At the graphene-water interface, $q$ remains nearly unchanged, despite local fluctuations in the first layer. In contrast, at the air-water interface, $q$ declines continuously to 0. When both interfaces have $N = 2.5$, $q$ is 0.69 at the graphene-water interface but only 0.56 at the air-water interface. These results indicate that hydrogen bonding remains tetrahedrally ordered at the graphene-water interface despite reduced coordination, whereas it weakens progressively to a gas-like state at the air-water interface.

The high-frequency positive peak in the HD-SFG signals at both interfaces, associated with free O-H groups, is commonly used to assess interfacial hydrophobicity[8], as similar features are observed at other hydrophobic interfaces, such as oil-water[45]. However, the intrinsic wettability of monolayer graphene remains debated based on different measurements and wettability metrics[46]. Since hydrophobic surfaces could enhance water diffusion and hydrophilic ones suppress it[47], here we further examine this feature by considering surface water dynamics. Figure 3(c) presents the mean-squared displacement ($MSD_{xy}$) of water along the $xy$ direction at equilibrium for three regions: the graphene-water interface (the first two layers), the air-water interface (the first two layers), and the bulk. To reduce the impact of molecular migration between regions, the nanosecond-scale simulation was segmented into 10 ps intervals, during which diffusion was computed for water molecules restricted to each region and subsequently ensemble-averaged. Both interfaces exhibit higher water diffusion coefficients than the bulk, with $D_{xy}^{bulk} = 2.43 \times 10^{-9}\, m^2/s$, $D_{xy}^{graphene} = 3.18 \times 10^{-9}\, m^2/s$, and $D_{xy}^{air} = 4.68 \times 10^{-9}\, m^2/s$, consistent with previous reports on bulk water diffusion[48], graphene-confined water[36], and the air-water interface[49]. The enhanced interfacial diffusion is attributed to the reduced structural correlation at the interface, increasing entropy and promoting molecular diffusion[50], in agreement with the nearest-neighbor oxygen variation in Figure 3b. Notably, water at the air-water interface exhibits diffusion behavior similar to ideal hydrophobic confinement (Fig. 3c)[51], serving as a reference for assessing surface hydrophobicity. The lower diffusion at the graphene-water interface compared to the air-water interface, along with the redshift of the high-frequency SFG peak, are consistent indicators. Despite this, the overall interfacial dynamics at the graphene-water interface are still enhanced. Combined with the presence of the free OH feature in HD-SFG signals, these findings suggest that monolayer graphene exhibits microscopic hydrophobic properties like air.

**CONCLUSIONS**

This study establishes a unified deep-learning-enabled framework for computing HD-SFG spectra with fully first-principles accuracy, with direct applications to the graphene–water and air–water interfaces. The potential energy surface, dipole, and polarizability models well capture the spectral features and physical properties at both the interface and in bulk, with the right qualitative trends and in good agreement with experimental results [9,39–41,48]. Our results reveal that enhanced water diffusivity and the emergence of a high-frequency positive peak in the HD-SFG response are consistent signatures of microscopic interfacial hydrophobicity, shared by both the solid–liquid and gas–liquid systems.

Despite their spectral resemblance, the microscopic behavior of interfacial water differs substantially between the two systems. At the graphene–water boundary, water forms a compact, non-centrosymmetric hydrogen-bond network constrained within a narrow interfacial region. In contrast, the air–water interface exhibits a broader transition zone, across which water gradually loses its tetrahedral coordination as it approaches the vapor phase. These differences, arising from the contrasting boundary conditions of solid–liquid and gas–liquid systems, underpin the variations in spectral decomposition observed in the HD-SFG response at the two interfaces.

Taken together, our findings disentangle the similarity of interfacial vibrational responses and expose the nuanced role of interfacial geometry and confinement in shaping molecular-level water behaviors. This work not only clarifies key spectroscopic observations but also paves the way for predictive, specific modeling of water at complex hydrophobic interfaces, with implications for materials design, surface science, and biological function.

# METHODS

## Density functional theory (DFT) calculations

To generate the initial training dataset for the potential energy surface (PES), *ab initio* molecular dynamics (AIMD) simulations at 340 K were conducted using the strongly constrained and appropriately normed (SCAN) exchange-correlation functional[35], as implemented in the Vienna *Ab initio* Simulation Package (VASP)[52]. The AIMD simulations employed a time step of 0.5 fs and were performed for a total duration of 12 ps. The projector-augmented wave (PAW) method[53], with a cutoff energy of 600 eV, was used to account for core-valence interactions and a total energy convergence criterion of $10^{-6}$ eV was imposed. The same computational settings were applied to static self-consistent DFT calculations throughout the active learning process[54] in constructing the machine learning model for the potential energy surface.

For the development of machine learning models that predict dipole moments and polarizability, additional DFT calculations using SCAN were carried out using configurations sampled from the potential energy surface dataset (including the active learning part). These calculations were performed with Quantum ESPRESSO[55] and Wannier90[56] packages to obtain maximally localized Wannier functions (MLWFs)[57] and compute the electronic polarizability at fixed nuclear positions. Polarizability was determined by numerically calculating the dipole change under a finite electric field of $\pm 0.001$ a.u., falling well within the linear response regime (Supplementary Fig. 1). Valence electron-ion interactions were described using Optimized Norm-Conserving Vanderbilt (ONCV) pseudopotentials, while electron wavefunctions were expanded in a plane-wave basis with a cutoff energy of 140 Ry. A total energy convergence threshold of $10^{-6}$ Ry was enforced.

## Deep neural network (DNN) models

The complete training data set for the deep learning models was constructed using an active learning framework[54], initialized from the data obtained through AIMD simulations. The data set for the PES model was curated from DFT calculations and comprises 920 graphene–water–air slab configurations, and 870 bulk water supercells, each containing 64 water molecules under ambient pressure. The explored temperature range spans from 270 K to 350 K.

For dipole moment and polarizability calculations, given the substantial computational cost of DFT calculations under an applied electric field, a subset of independent graphene–water–air slab configurations including the active learning part was selected to compute Wannier centers (WCs) and polarizability in the presence of an electric field. The Deep Potential[23], Deep Wannier[27], and Deep Polarizability[28] models were trained on this dataset using the DeePMD-kit package[58].

**MLMD simulations**

Machine learning molecular dynamics (MLMD) simulations were performed using the LAMMPS[59] package to sample molecular configurations. The MLMD supercell consists of 60 carbon atoms forming a graphene monolayer and 166 water molecules, with dimensions of 12.01 Å ×12.78 Å in the $xy$-plane and 50 Å along the $z$-axis, including a ~15 Å vacuum layer to mitigate periodic boundary effects[8]. The MLMD simulations were conducted in the NVT ensemble using a Nosé–Hoover thermostat[60,61]. The temperature damping parameter was set to 0.5 ps, and a time step of 0.5 fs was employed to integrate the equations of motion. To enhance molecular sampling and minimize statistical errors, twelve independent MLMD simulations were performed at 340 K for 1 ns, with the first 0.1 ns dedicated to equilibration. A 40 K temperature elevation was applied to mitigate the SCAN functional's tendency to overestimate the melting point of ice[37], as most of the sum-frequency generation (SFG) spectra were calculated within the 300 K–330 K range[21,43]. The molecular dipole moment and polarizability were computed on-the-fly using trained models with the DeePMD-kit package[58], alongside molecular configurations sampled from the MLMD trajectories. Detailed post-calculations and analyses—such as SFG spectral computations and diffusion analysis based on the MLMD trajectories—are provided in the Supplementary Information.

**REFERENCES**

1. Verdaguer, A., Sacha, G. M., Bluhm, H. & Salmeron, M. Molecular Structure of Water at Interfaces: Wetting at the Nanometer Scale. *Chem. Rev.* **106**, 1478–1510 (2006).
2. Donaldson, D. J. & Vaida, V. The Influence of Organic Films at the Air−Aqueous Boundary on Atmospheric Processes. *Chem. Rev.* **106**, 1445–1461 (2006).
3. Xu, K., Cao, P. & Heath, J. R. Graphene Visualizes the First Water Adlayers on Mica at Ambient Conditions. *Science* **329**, 1188–1191 (2010).
4. Zou, X. *et al.* Molecular Interactions between Graphene and Biological Molecules. *J. Am. Chem. Soc.* **139**, 1928–1936 (2017).
5. Algara-Siller, G. *et al.* Square ice in graphene nanocapillaries. *Nature* **519**, 443–445 (2015).

6. Kapil, V. *et al.* The first-principles phase diagram of monolayer nanoconfined water. *Nature* **609**, 512–516 (2022).
7. Shen, Y. R. & Ostroverkhov, V. Sum-Frequency Vibrational Spectroscopy on Water Interfaces: Polar Orientation of Water Molecules at Interfaces. *Chem. Rev.* **106**, 1140–1154 (2006).
8. Tang, F. *et al.* Molecular Structure and Modeling of Water–Air and Ice–Air Interfaces Monitored by Sum-Frequency Generation. *Chem. Rev.* **120**, 3633–3667 (2020).
9. Wang, Y. *et al.* Heterodyne-Detected Sum-Frequency Generation Vibrational Spectroscopy Reveals Aqueous Molecular Structure at the Suspended Graphene/Water Interface. *Angew. Chem. Int. Ed.* **63**, e202319503 (2024).
10. Xu, Y., Ma, Y.-B., Gu, F., Yang, S.-S. & Tian, C.-S. Structure evolution at the gate-tunable suspended graphene–water interface. *Nature* **621**, 506–510 (2023).
11. Kim, D. *et al.* Wettability of graphene and interfacial water structure. *Chem* **7**, 1602–1614 (2021).
12. Montenegro, A. *et al.* Asymmetric response of interfacial water to applied electric fields. *Nature* **594**, 62–65 (2021).
13. Calegari Andrade, M. F., Ko, H.-Y., Car, R. & Selloni, A. Structure, Polarization, and Sum Frequency Generation Spectrum of Interfacial Water on Anatase $TiO_2$. *J. Phys. Chem. Lett.* **9**, 6716–6721 (2018).
14. Ohto, T., Tada, H. & Nagata, Y. Structure and dynamics of water at water–graphene and water–hexagonal boron-nitride sheet interfaces revealed by *ab initio* sum-frequency generation spectroscopy. *Phys. Chem. Chem. Phys.* **20**, 12979–12985 (2018).
15. Ohto, T., Usui, K., Hasegawa, T., Bonn, M. & Nagata, Y. Toward *ab initio* molecular dynamics modeling for sum-frequency generation spectra; an efficient algorithm based on surface-specific velocity-velocity correlation function. *J. Chem. Phys.* **143**, 124702 (2015).
16. Olivieri, J.-F., Hynes, J. T. & Laage, D. Water dynamics and sum-frequency generation spectra at electrode/aqueous electrolyte interfaces. *Faraday Discuss.* **249**, 289–302 (2024).
17. Car, R. & Parrinello, M. Unified Approach for Molecular Dynamics and Density-Functional Theory. *Phys. Rev. Lett.* **55**, 2471–2474 (1985).
18. Chen, M. *et al.* Ab initio theory and modeling of water. *Proc. Natl. Acad. Sci.* **114**, 10846–10851 (2017).
19. Groß, A. & Sakong, S. Ab Initio Simulations of Water/Metal Interfaces. *Chem. Rev.* **122**, 10746–10776 (2022).
20. Nagata, Y. & Mukamel, S. Vibrational Sum-Frequency Generation Spectroscopy at the Water/Lipid Interface: Molecular Dynamics Simulation Study. *J. Am. Chem. Soc.* **132**, 6434–6442 (2010).
21. Du, X. *et al.* Revealing the molecular structures of $\alpha$-$Al_2O_3$(0001)–water interface by machine learning based computational vibrational spectroscopy. *J. Chem. Phys.* **161**, 124702 (2024).
22. Behler, J. & Parrinello, M. Generalized Neural-Network Representation of High-Dimensional Potential-Energy Surfaces. *Phys. Rev. Lett.* **98**, 146401 (2007).
23. Zhang, L., Han, J., Wang, H., Car, R. & E, W. Deep Potential Molecular Dynamics: A Scalable Model with the Accuracy of Quantum Mechanics. *Phys. Rev. Lett.* **120**, 143001 (2018).
24. Fan, Z. *et al.* GPUMD: A package for constructing accurate machine-learned potentials and performing highly efficient atomistic simulations. *J. Chem. Phys.* **157**, 114801 (2022).

25. Wang, J. *et al.* E(n)-Equivariant cartesian tensor message passing interatomic potential. *Nat. Commun.* **15**, 7607 (2024).
26. Litman, Y., Chiang, K.-Y., Seki, T., Nagata, Y. & Bonn, M. Surface stratification determines the interfacial water structure of simple electrolyte solutions. *Nat. Chem.* **16**, 644–650 (2024).
27. Zhang, L. *et al.* Deep neural network for the dielectric response of insulators. *Phys. Rev. B* **102**, 041121 (2020).
28. Sommers, G. M., Calegari Andrade, M. F., Zhang, L., Wang, H. & Car, R. Raman spectrum and polarizability of liquid water from deep neural networks. *Phys. Chem. Chem. Phys.* **22**, 10592–10602 (2020).
29. Schienbein, P. Spectroscopy from Machine Learning by Accurately Representing the Atomic Polar Tensor. *J. Chem. Theory Comput.* **19**, 705–712 (2023).
30. Litman, Y., Lan, J., Nagata, Y. & Wilkins, D. M. Fully First-Principles Surface Spectroscopy with Machine Learning. *J. Phys. Chem. Lett.* **14**, 8175–8182 (2023).
31. De La Puente, M., Gomez, A. & Laage, D. Neural Network-Based Sum-Frequency Generation Spectra of Pure and Acidified Water Interfaces with Air. *J. Phys. Chem. Lett.* **15**, 3096–3102 (2024).
32. Kapil, V., Kovács, D. P., Csányi, G. & Michaelides, A. First-principles spectroscopy of aqueous interfaces using machine-learned electronic and quantum nuclear effects. *Faraday Discuss.* **249**, 50–68 (2024).
33. Berrens, M. L., Calegari Andrade, M. F., Fourkas, J. T., Pham, T. A. & Donadio, D. Molecular Fingerprints of Ice Surfaces in Sum Frequency Generation Spectra: A First-Principles Machine Learning Study. *JACS Au* jacsau.4c00957 (2025) doi:10.1021/jacsau.4c00957.
34. Rashmi, R. *et al.* Revealing the Water Structure at Neutral and Charged Graphene/Water Interfaces through Quantum Simulations of Sum Frequency Generation Spectra. *ACS Nano* **19**, 4876–4886 (2025).
35. Sun, J., Ruzsinszky, A. & Perdew, J. P. Strongly Constrained and Appropriately Normed Semilocal Density Functional. *Phys Rev Lett* **115**, 036402 (2015).
36. Cicero, G., Grossman, J. C., Schwegler, E., Gygi, F. & Galli, G. Water Confined in Nanotubes and between Graphene Sheets: A First Principle Study. *J. Am. Chem. Soc.* **130**, 1871–1878 (2008).
37. Piaggi, P. M., Panagiotopoulos, A. Z., Debenedetti, P. G. & Car, R. Phase Equilibrium of Water with Hexagonal and Cubic Ice Using the SCAN Functional. *J. Chem. Theory Comput.* **17**, 3065–3077 (2021).
38. Zhang, C. *et al.* Modeling Liquid Water by Climbing up Jacob's Ladder in Density Functional Theory Facilitated by Using Deep Neural Network Potentials. *J. Phys. Chem. B* **125**, 11444–11456 (2021).
39. Birnbaum, G. & Chatterjee, S. K. The Dielectric Constant of Water Vapor in the Microwave Region. *J. Appl. Phys.* **23**, 220–223 (1952).
40. Badyal, Y. S. *et al.* Electron distribution in water. *J. Chem. Phys.* **112**, 9206–9208 (2000).
41. Murphy, W. F. The Rayleigh depolarization ratio and rotational Raman spectrum of water vapor and the polarizability components for the water molecule. *J. Chem. Phys.* **67**, 5877–5882 (1977).
42. Baer, M. D., Tobias, D. J. & Mundy, C. J. Investigation of Interfacial and Bulk Dissociation of HBr, HCl, and $HNO_3$ Using Density Functional Theory-Based Molecular Dynamics Simulations. *J. Phys. Chem. C* **118**, 29412–29420 (2014).

43. Ohto, T. *et al.* Accessing the Accuracy of Density Functional Theory through Structure and Dynamics of the Water–Air Interface. *J. Phys. Chem. Lett.* **10**, 4914–4919 (2019).
44. Paolantoni, M., Lago, N. F., Albertí, M. & Laganà, A. Tetrahedral Ordering in Water: Raman Profiles and Their Temperature Dependence. *J. Phys. Chem. A* **113**, 15100–15105 (2009).
45. Moore, F. G. & Richmond, G. L. Integration or Segregation: How Do Molecules Behave at Oil/Water Interfaces? *Acc. Chem. Res.* **41**, 739–748 (2008).
46. Kim, E. *et al.* Wettability of graphene, water contact angle, and interfacial water structure. *Chem* **8**, 1187–1200 (2022).
47. Agosta, L., Arismendi-Arrieta, D., Dzugutov, M. & Hermansson, K. Origin of the Hydrophobic Behaviour of Hydrophilic $CeO_2$. *Angew. Chem. Int. Ed.* **62**, e202303910 (2023).
48. Holz, M., Heil, S. R. & Sacco, A. Temperature-dependent self-diffusion coefficients of water and six selected molecular liquids for calibration in accurate 1H NMR PFG measurements. *Phys. Chem. Chem. Phys.* **2**, 4740–4742 (2000).
49. Liu, P., Harder, E. & Berne, B. J. Hydrogen-Bond Dynamics in the Air−Water Interface. *J. Phys. Chem. B* **109**, 2949–2955 (2005).
50. Agosta, L., Briels, W., Hermansson, K. & Dzugutov, M. The entropic origin of the enhancement of liquid diffusion close to a neutral confining surface. *J. Chem. Phys.* **161**, 091102 (2024).
51. Agosta, L., Hermansson, K. & Dzugutov, M. Water under hydrophobic confinement: entropy and diffusion. *ArXivcond-Matsoft* **2412.03726**, (2024).
52. Kresse, G. & Furthmüller, J. Efficient iterative schemes for *ab initio* total-energy calculations using a plane-wave basis set. *Phys. Rev. B* **54**, 11169–11186 (1996).
53. Kresse, G. & Joubert, D. From ultrasoft pseudopotentials to the projector augmented-wave method. *Phys. Rev. B* **59**, 1758–1775 (1999).
54. Zhang, Y. *et al.* DP-GEN: A concurrent learning platform for the generation of reliable deep learning based potential energy models. *Comput. Phys. Commun.* **253**, 107206 (2020).
55. Giannozzi, P. *et al.* Advanced capabilities for materials modelling with Quantum ESPRESSO. *J. Phys. Condens. Matter* **29**, 465901 (2017).
56. Pizzi, G. *et al.* Wannier90 as a community code: new features and applications. *J. Phys. Condens. Matter* **32**, 165902 (2020).
57. Marzari, N. & Vanderbilt, D. Maximally localized generalized Wannier functions for composite energy bands. *Phys. Rev. B* **56**, 12847–12865 (1997).
58. Wang, H., Zhang, L., Han, J. & E, W. DeePMD-kit: A deep learning package for many-body potential energy representation and molecular dynamics. *Comput. Phys. Commun.* **228**, 178–184 (2018).
59. Thompson, A. P. *et al.* LAMMPS - a flexible simulation tool for particle-based materials modeling at the atomic, meso, and continuum scales. *Comput. Phys. Commun.* **271**, 108171 (2022).
60. Nosé, S. A molecular dynamics method for simulations in the canonical ensemble. *Mol. Phys.* **100**, 191–198 (2002).
61. Hoover, W. G. Canonical dynamics: Equilibrium phase-space distributions. *Phys. Rev. A* **31**, 1695–1697 (1985).

**Acknowledgements**

We are grateful to Prof. Chuan-Shan Tian and Prof. Mischa Bonn for sharing their experimental data and for valuable discussions that improved our understanding of the results. We also thank Dr. Axel Gomez for his insightful comments. This work was conducted within the Computational Chemical Science Center: Chemistry in Solution and at Interfaces funded by the U.S. Department of Energy under Award No. DE-SC0019394, and supported by the Swedish Research Council (Vetenskapsrådet), grant number 201703950. The authors acknowledge that the work reported in this paper was performed using the Princeton Research Computing resources at Princeton University. This research also used resources from the National Energy Research Scientific Computing Center (NERSC) operated under Contract No. DE-AC02-05CH11231 using NERSC award ERCAP0021510.

**Author contributions**

A.S. and R.C. designed the project. Y.W. carried out the simulations and performed the analysis. Y.W., A.S. and R.C. wrote the manuscript. Y.L., L.A., and M.C.A. reviewed the manuscript. L.D., and C.Z. contributed fruitful discussions.

**Competing interests**

The authors declare no competing interests.

# Figures

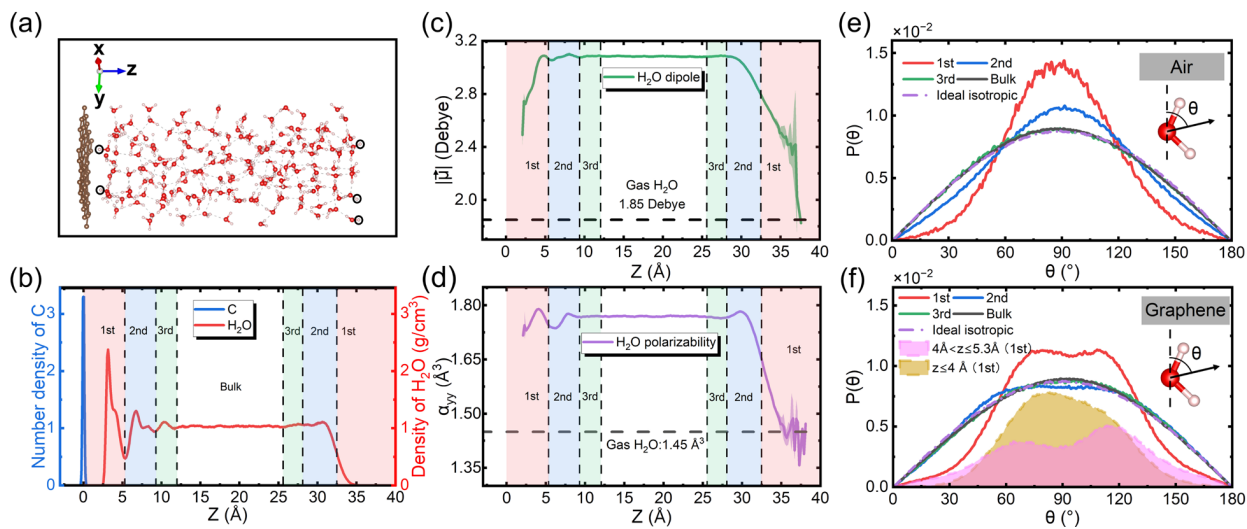

**Fig. 1** | (**a**) Side-view of the graphene-water-air slab in MLMD simulations, with black circles marking dangling OH bonds. (**b**) Number density profiles of carbon and water density along *z*. (**c, d**) ML-calculated water dipole moment and $\alpha_{yy}$ distributions along *z*, aligned with the density profile. Dashed lines indicate experimental gas-phase values[39,41]. (**e, f**) Normalized probability distributions of the dipole angle ($\theta$) relative to the surface normal at the air-water (e) and graphene-water (f) interfaces across layers. In (f), shaded regions highlight first-layer subregions near and away from graphene.

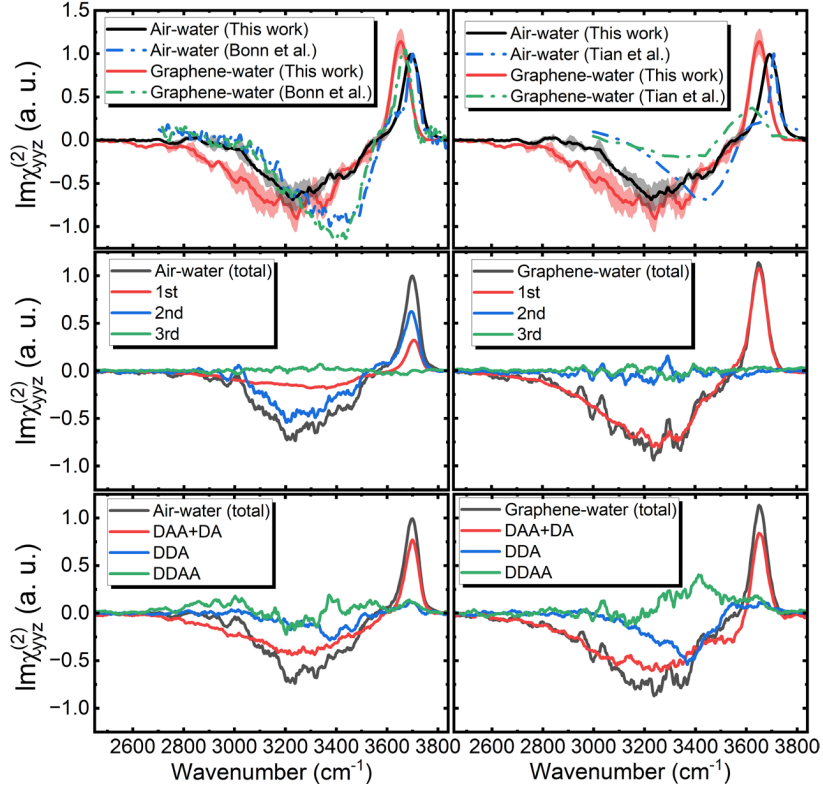

**Fig. 2** | The calculated $Im\chi_{yyz}(\omega)$ spectra are compared with results from two substrate-free experiments: Ref.[9] (E1, top left) and Ref.[10] (E2, top right). For the decomposition of the calculated $Im\chi^{(2)}_{yyz}(\omega)$ spectra, the contributions of the three interfacial layers defined in Fig. 1b are shown (in the middle panels), while the decomposition into water species with distinct H-bond characteristics at the interfaces is presented (in the bottom panels).

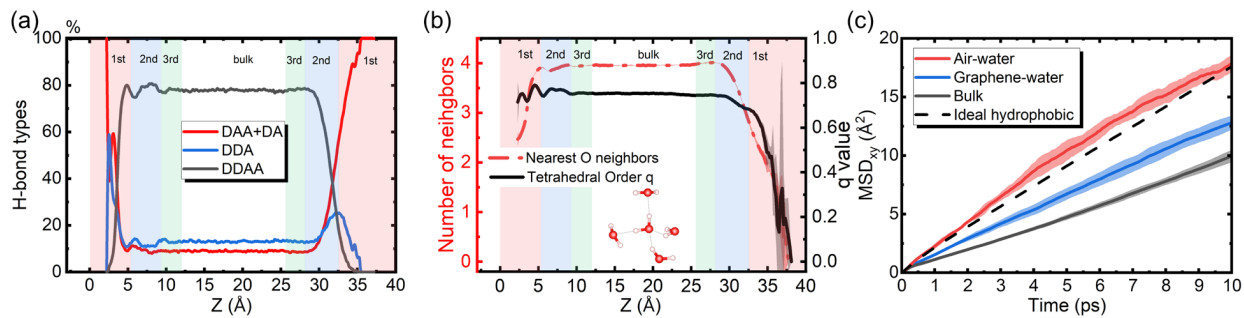

**Fig. 3** | **(a)** Proportions of hydrogen-bond groups along the $z$-axis, categorized by the number of donor and acceptor sites for each water molecule. **(b)** $z$-axis profiles of the nearest-neighbor count for water molecules and their corresponding orientational tetrahedral order $q$. The inset depicts the tetrahedral hydrogen-bond network in the bulk region obtained from the MLMD simulation, corresponding to the DDAA grouping in (a). **(c)** Mean-squared displacements of water along the $xy$ direction (parallel to the interface) in different regions of the graphene-water-air slab and at the ideal hydrophobic wall.